\documentclass[11pt,twoside]{article}
\usepackage{asp2010}

\resetcounters

\begin{document}

\title{The internal rotation of the GW Vir star PG0112$+$200 
through the eyes of asteroseismology} 

\author{A. H. C\'orsico$^1$, 
        L. G. Althaus$^1$, 
        S. D. Kawaler$^2$,
        M. M. Miller Bertolami$^1$, and 
        E. Garc\'\i a--Berro$^{3,4}$
\affil{$^1$Facultad de Ciencias Astron\'omicas y Geof\'isicas, 
           Universidad Nacional de La Plata, 
           Paseo del Bosque s/n, 
           (1900) La Plata, 
           Argentina}
\affil{$^2$Department of Physics and Astronomy, 
           Iowa State University,
           12 Physics Hall, Ames, 
           IA 50011, 
           U.S.A.}
\affil{$^3$Departament de F\'\i sica Aplicada, 
           Universitat Polit\`ecnica de Catalunya,
           c/Esteve Terrades 5, 
           08860 Castelldefels, 
           Spain}
\affil{$^4$Institute for Space Studies of Catalonia, 
           c/Gran Capit\`a 2-4, Edif. Nexus 104, 
           08034 Barcelona, 
           Spain}}

\begin{abstract}
We   investigate  the   internal  rotation   profile  of   the  GW~Vir
(PG1159-type) star  PG~0122$+$200 by employing  an asteroseismological
model that closely  reproduces the observed periods of  this star.  We
adopt  a forward  approach  and  two inversion  methods  based on  the
rotational  splitting  of the  pulsation  frequencies  to explore  the
properties of  the rotation of  PG~0122$+$200.  We found  evidence for
differential rotation in this star.
\end{abstract}

\section{Introduction}

In  recent  years, asteroseismology  has  become  a  powerful tool  to
unravel the  internal structure  of oscillating stars,  in particular,
pulsating white  dwarfs and pre-white  dwarfs (Winget \&  Kepler 2008,
Althaus  et  al.  2010).   The  approach is  suitable  also  to  place
constraints on the stellar rotation. In fact, stellar rotation removes
the intrinsic mode degeneracy of a nonradial $g$-mode characterized by
an harmonic degree  $\ell$ and a radial order $k$.   As a result, each
pulsation   frequency  is   split  into   multiplets   of  $(2\ell+1)$
frequencies specified by different  values of the azimuthal index $m$,
with  $m= 0, \pm  1, \ldots,  \pm \ell$  (Unno et  al.  1989).  If the
rotation  rate is slow  compared with  the pulsation  frequencies, the
frequency  separation  between  each  component of  the  multiplet  is
proportional to the rotation velocity of the star ($\Omega$).  In this
work, we perform a detailed asteroseismological study aimed at placing
constraints on the internal rotation of the GW Vir star PG 0122$+$200,
using the best existing evolutionary and seismic model for this star.

\section{PG 0122$+$200 and the asteroseismological model}

PG 0122$+$200 has  $T_{\rm eff}= 80\, 000 \pm 4\,  000$~K and $\log g=
7.5\pm  0.5$ (Dreizler  \& Heber  1998).  This  pulsating  PG1159 star
currently  defines the  locus of  the low-luminosity  red edge  of the
GW~Vir instability  strip (Althaus et  al.  2010).  In this  paper, we
employ the  high-quality observational data on  PG 0122$+$200 gathered
by Fu  et al.  (2007),  consisting of 23 frequencies  corresponding to
modes  with $\ell=  1$  that include  seven  rotational triplets  ($m=
-1,0,+1$) and two isolated frequencies  with (probably) $m= 0$ --- see
Table 5  of Fu et  al.  (2007).  The theoretical  frequency splittings
were assessed employing  the non-rotating asteroseismological model of
PG  0122$+$200  derived  by  C\'orsico  et al.   (2007).   This  model
reproduces the $m=  0$ observed periods with an  average of the period
differences   (theoretical  vs.    observed)   of  $\lesssim   0.9$~s.
Employing  an  asteroseismological   model  represents  a  substantial
improvement over previous works of  this kind --- see, e.g., Charpinet
et al. (2009), for the case of PG~1159$-$035.

\section{The forward approach}

This method has  been described in detail in  Charpinet et al. (2009).
The theoretical frequency splittings  are obtained varying the assumed
rotation profile  and then  they are compared  with the  observed ones
until a  best global match  is found.  The theoretical  splittings are
computed using the expressions  resulting from the perturbative theory
to  first order  in $\Omega$,  that  assumes that  the pulsating  star
rotates with  a period much longer  than any of  its pulsation periods
(Unno et  al.  1989.   The goodness of  the match  between theoretical
($\delta  \nu_i^{\rm  T}$)   and  observed  ($\delta  \nu_i^{\rm  O}$)
rotational splittings  is described  using a quality  function defined
as:
$$\chi^2=\frac{1}{N_{\rm s}}\sum_{i=1}^{N_{\rm s}}
\frac{1}{\sigma_i^{2}}(\delta\nu_{i}^{\rm  T}  -  
\delta  \nu_{i}^{\rm O})^2.$$ 
Each difference is  weighted with the inverse of  the squared standard
uncertainty ($\sigma_i$) of the  observed splittings, which is derived
from the uncertainties in the frequencies given in Fu et al.  (2007).

We  used  very  simple   functional  forms  for  $\Omega(r)$,  and  in
particular,      linear      differential      rotation      profiles,
$\Omega(r)=(\Omega_{\rm s}-\Omega_{\rm c})\ r + \Omega_{\rm c}$, where
$\Omega_{\rm s}$  and $\Omega_{\rm c}$  are the rotation rates  at the
stellar  surface  and  center,  respectively.   We  performed  forward
computations varying the  parameters $\Omega_{\rm s}$ and $\Omega_{\rm
c}$ in the range $0-20$~$\mu$Hz.  We found that there exists a unique,
well-localized  best-fit solution at  $\Omega_{\rm c}=  10.62\pm 1.8\,
\mu$Hz  and  $\Omega_{\rm  s})=  4.41\pm  1.1\,  \mu$Hz.   The  quoted
uncertainties are  derived by means of  an error analysis  in which we
assume  that the  uncertainties of  the observed  frequency splittings
have a Gaussian distribution with  a standard deviation of $\sim 0.08$
$\mu$Hz.   The existence of  this solution  suggests that  the central
regions of PG 0122$+$200 could be rotating more than twice faster than
the  surface.   These results  disagree  with  those  of C\'orsico  \&
Althaus (2010), that indicate  that the surface of PG~0122$+$200 could
be spinning faster  than the core.  The discrepancy  has its origin in
the fact  that, in that preliminary  work, the fits  of the rotational
splittings were  made without  weighting the terms  of the sum  in the
quality function, and thus,  the impact of the different uncertainties
of the observational data on the final result was neglected.

\begin{figure*} 
\begin{center}
\includegraphics[clip,width=13 cm]{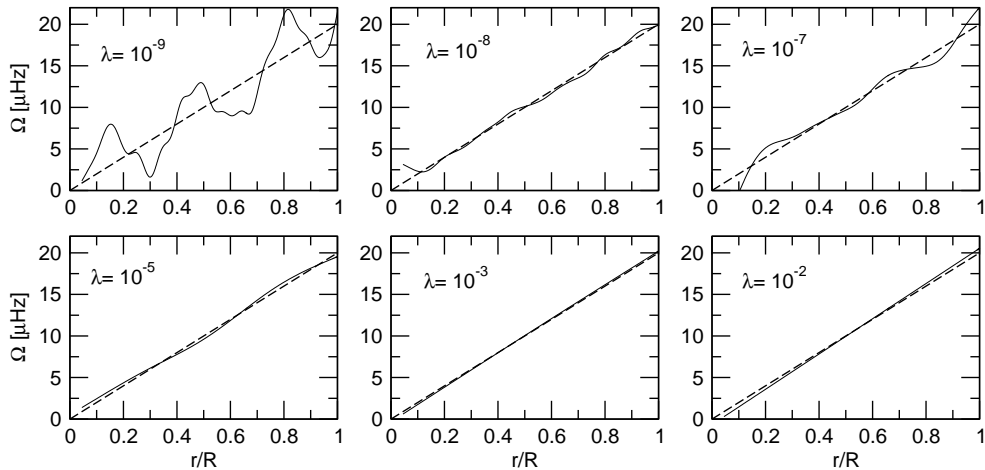} 
\caption{A test of our RLS  method. The dashed straight line shows the
  input   rotation   profile   ($\Omega_{\rm  c}=   10^{-4}\   \mu$Hz,
  $\Omega_{\rm c}=  20\ \mu$Hz), and the thin  solid curve corresponds
  to  the inverted  profile  for increasing  values  of the  parameter
  $\lambda$.}
\label{figure1} 
\end{center}
\end{figure*} 

Vauclair et  al. (2011) have found that  seven oscillation frequencies
of PG~0122$+$200 are experiencing  changes over time, with much larger
amplitudes and shorter time scales  than those expected by cooling. We
have redone  our forward  computations by considering  these frequency
drifts.  We  have  arrived  at  the conclusion  that,  even  with  the
uncertainties  produced   by  the  changes  of   some  frequencies  of
PG~0122$+$200 over time, rigid rotation can be discarded.

\section{Other inversion methods}

\begin{figure*} 
\begin{center}
\includegraphics[clip,width=10 cm]{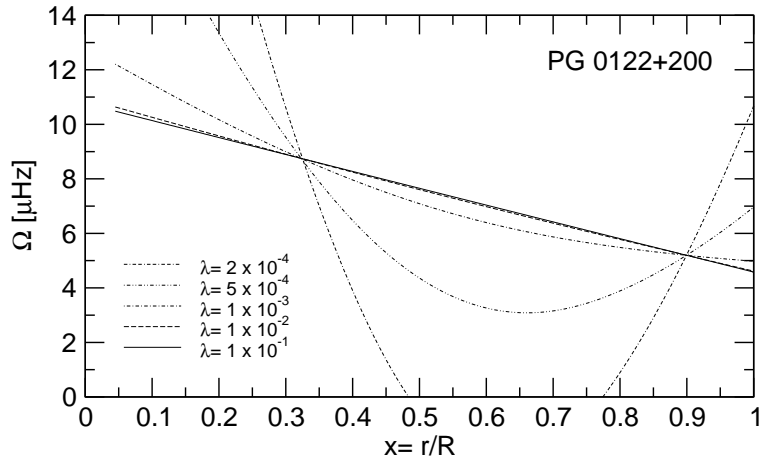} 
\caption{Inverted rotation profiles for PG~0122$+$200 corresponding to
  different values of the regularization parameter $\lambda$.}
\label{figure2} 
\end{center}
\end{figure*}

We  have  gone  beyond  the  forward approach,  and  employed  another
different,  independent  method to  explore  the  rotation profile  of
PG~0122$+$200.  This is  the Regularized  Least Squares  (RLS) fitting
technique --- see Kawaler et al.  (1999) for details.  We first tested
the  reliability of  our RLS  scheme  by employing  this technique  on
``synthetic''     frequency    splittings    generated     with    the
asteroseismological  model   of  PG~0122$+$200  through   the  forward
approach.    Specifically,   we   considered   rotational   splittings
corresponding to  consecutive $\ell=  1$ $g$-modes with  $k= 1,\cdots,
40$.  The  regularization matrix corresponded to the  smoothing of the
second derivative  of $\Omega(r)$.   In all of  the cases  examined so
far, the inversions were able to recover the input rotation profile we
used to  compute the synthetic  splittings, provided that  an adequate
range of values of the regularization parameter $\lambda$ was adopted.
Fig.~\ref{figure1}  illustrates  the  particular  case of  a  linearly
increasing rotation profile.  Next, we applied the RLS method to infer
the internal rotation profile  of PG~0122$+$200.  We have employed the
seven observed (averaged) $\ell= 1$ splittings.  In Fig.~\ref{figure2}
we  show the  inverted rotation  profiles.  For  very small  values of
$\lambda$, the  inverted profiles exhibit strong  variations that lack
physical meaning.   However, as the  value of $\lambda$  is increased,
the  inverted solution gradually  stabilizes.  The  resulting rotation
profile (corresponding  to $\lambda  \gtrsim 10^{-2}$) consists  of an
almost  linearly decreasing  rotation rate  with $\Omega_{\rm  c} \sim
10.75\, \mu$Hz  and $\Omega_{\rm s}  \sim 4.58\, \mu$Hz,  in excellent
agreement with  the results  of the forward  approach. An  analysis of
errors similar to the one  performed for the forward approach leads to
the   conclusion  that   even   with  the   inclusion  of   reasonable
uncertainties   in   the   observed   splittings,  the   rotation   of
PG~0122$+$200  is   faster  at  the   central  regions  than   at  the
surface. Specifically, we found $\Omega_{\rm c}= 10.75\pm 2.4\, \mu$Hz
and $\Omega_{\rm s}= 4.58\pm 1.7\, \mu$Hz.

Finally,  we  have  also  made  rotational  inversions  onto  a  fixed
functional basis.   The method is called ``function  fitting'' --- for
details, see Kawaler et  al.  (1999). Specifically, we explored linear
functional  forms for $\Omega(r)$.   The optimal  values we  found are
$\Omega_{\rm c}=  10.74\pm 2.9\,  \mu$Hz and $\Omega_{\rm  s}= 4.57\pm
1.8\, \mu$Hz, in  excellent agreement with the RLS  fits and also with
the forward approach.

\section{Conclusions}

The  three methods  employed  in  this work  to  explore the  internal
rotation of  PG~0122$+$200 suggest  that it is differential,  with the
central  regions   rotating  more  than  twice   faster  than  stellar
surface.  This   constitutes  the  first   asteroseismic  evidence  of
differential rotation with depth for an evolved star.

\acknowledgements 

One  of us  (A.H.C.)   warmly thanks  Prof.   Hiromoto Shibahashi  for
support, that allowed him to attend the conference.

{}

\end{document}